\documentclass[prd,preprint,nofootinbib]{revtex4}
\usepackage{amssymb}
\usepackage{epsfig}

\begin{document}

\title{ 
Relic density of dark matter in brane world cosmology
}

\author{Nobuchika Okada}
 \email{okadan@post.kek.jp}
 \affiliation{
  Theory Division, KEK, Oho 1-1, Tsukuba, Ibaraki 305-0801, Japan
 }

\author{Osamu Seto}
 \email{osamu@mail.nctu.edu.tw}
 \affiliation{
 Institute of Physics, National Chiao Tung University, 
 Hsinchu, Taiwan 300, Republic of China
}

\begin{abstract}
We investigate the thermal relic density of the cold dark matter
 in the context of the brane world cosmology.
Since the expansion law in a high energy regime is modified
 from the one in the standard cosmology,
 if the dark matter decouples in such a high energy regime  
 its relic number density is affected by this modified expansion law.
We derive analytic formulas for the number density
 of the dark matter.
It is found that the resultant relic density is characterized
 by the ``transition temperature'' at which
 the modified expansion law in the brane world cosmology
 is connecting with the standard one,
 and can be considerably enhanced
 compared to that in the standard cosmology,
 if the transition temperature is low enough.
\end{abstract}

\preprint{KEK-TH-969}

\vspace*{3cm}
\maketitle


Recent various cosmological observations,
 especially WMAP satellite \cite{WMAP},
 have established the $\Lambda$CDM cosmological model
 with a great accuracy,
 where the energy density in the present universe
 consists of about $73\%$ of the cosmological constant (dark energy),
 $23\%$ of non-baryonic cold dark matter and just $4\%$ of baryons.
However, to clarify the identity of the dark matter particle is still
 a prime open problem in cosmology and particle physics.
Many candidates for dark matter have been proposed.
Among them, the neutralino in supersymmetric models
 is a suitable candidate,
 if the neutralino is the lightest supersymmetric particle (LSP)
 and the R-parity is conserved \cite{DMreview}.

In the case that the dark matter is the thermal relic,
 we can estimate its number density by solving
 the Boltzmann equation \cite{Kolb},
\begin{equation}
\frac{d n}{d t}+3Hn = -\langle\sigma v\rangle(n^2-n_{EQ}^2),
\label{n;Boltzmann}
\end{equation}
with the Friedmann equation,
\begin{equation}
H^2 = \frac{8\pi G}{3}\rho,
\label{FriedmannEq}
\end{equation}
where $H \equiv\dot{a}/a$ is the Hubble parameter
 with $a(t)$ being the scale factor,
 $n$ is the actual number density,
 $n_{EQ}$ is the number density in thermal equilibrium,
 $\langle\sigma v\rangle$ is the thermal averaged product
 of the annihilation cross section $\sigma$ and the relative velocity $v$,
 $\rho$ is the energy density,
 and $G$ is the Newton's gravitational constant.
By using $\dot{\rho}/\rho = -4H = 4\dot{T}/T+\dot{g_*}/g_*$,
 where $g_*$ is the effective total number
 of relativistic degrees of freedom,
 in terms of the number density to entropy ratio $Y = n/s$
 and $x = m/T$,
 Eq. (\ref{n;Boltzmann}) can be rewritten as
\begin{eqnarray}
\frac{d Y}{d x}
= -\frac{s\langle\sigma v\rangle}{xH}(Y^2-Y_{EQ}^2) ,
\label{Y;Boltzmann}
\end{eqnarray}
 if $\dot{g_*}/g_*$ is almost negligible as usual.
As is well known, an approximate formula of the solution
 of the Boltzmann equation can be described as
\begin{equation}
Y(\infty) \simeq \frac{x_d}
{\lambda \left(\sigma_0 +\frac{1}{2} \sigma_1 x_d^{-1} \right)},
\label{Ystandard}
\end{equation}
with a constant $\lambda = xs/H = 0.26 (g_{*S}/g_*^{1/2}) M_{P} m$
 for models in which $\langle\sigma v\rangle$ is approximately
 parameterized as
 $\langle \sigma v\rangle = \sigma_0  + \sigma_1 x^{-1}
 +\mathcal{O}(x^{-2})$,
 where $x_d = m/T_d$, $T_d$ is the decoupling temperature
 and $m$ is the mass of the dark matter particle,  
 and $M_P \simeq 1.2 \times 10^{19}$GeV is the Planck mass.

Recently, the brane world models have been attracting
 a lot of attention as a novel higher dimensional theory.
In these models, it is assumed that the standard model particles
 are confined on a ``3-brane'' while gravity resides in
 the whole higher dimensional spacetime.
The model first proposed by Randall and Sundrum (RS) \cite{RS},
 the so-called RS II model, is a simple and interesting one,
 and its cosmological evolution have been
 intensively investigated \cite{braneworld}.
In the model, our 4-dimensional universe is realized
 on the 3-brane with a positive tension
 located at the ultra-violet boundary
 of a five dimensional Anti de-Sitter spacetime.
In this setup, the Friedmann equation for a spatially flat spacetime
 in the RS brane cosmology is found to be
\begin{equation}
H^2 = \frac{8\pi G}{3}\rho\left(1+\frac{\rho}{\rho_0}\right) +
\frac{C}{a^4},
\label{BraneFriedmannEq}
\end{equation}
where
\begin{eqnarray}
\rho_0 = 96 \pi G M_5^6,
\end{eqnarray}
with $M_5$ being the five dimensional Planck mass,
 the third term with a integration constant $C$
 is referred to as the ``dark radiation'',
 and we have omitted the four dimensional cosmological constant.
The second term proportional to $\rho^2$ and the dark radiation
 are new ingredients in the brane world cosmology
 and lead to a non-standard expansion law.

In this paper, we investigate the brane cosmological effect
 for the relic density of the dark matter
 due to this non-standard expansion law.
If the new terms in Eq.~(\ref{BraneFriedmannEq}) dominates
 over the term in the standard cosmology
 at the freeze out time of the dark matter,
 they can cause a considerable modification
 for the relic abundance of the dark matter
 as we will show later.  

Before turning to our analysis, we give some comments here.
First, the dark radiation term is severely constrained
 by the success of the Big Bang Nucleosynthesis (BBN),
 since the term behaves like an additional radiation
 at the BBN era \cite{ichiki}.
Hence, for simplicity, we neglect the term in the following analysis.
Even if we include non-zero $C$ consistent with the BBN constraint,
 we cannot expect the significant effects from the dark radiation
 since at the era we will discuss the contribution
 of the dark radiation is negligible.
The second term is also constrained by the BBN,
 which is roughly estimated as $\rho_0^{1/4} > 1$ MeV
 (or $M_5 > 8.8$ TeV).
On the other hand, more sever constraint is obtained
 by the precision measurements of the gravitational law
 in sub-millimeter range.
Through the vanishing cosmological constant condition,
 we find $\rho_0^{1/4} > 1.3$ TeV (or $M_5 > 1.1 \times 10^8$ GeV)
 discussed in the original paper by Randall and Sundrum \cite{RS}.
However note that this result, in general, is quite model dependent.
For example, if we consider an extension of the model
 so as to introduce a bulk scalar field,
 the constraint can be moderated as discussed in \cite{maedawands}.
Hence, hereafter we impose only the BBN constraint on $\rho_0$.

We are interested in the early stage in the brane world cosmology
 where the $\rho^2$ term dominates, namely, $\rho^2/\rho_0 \gg \rho$.
In this case the coupling factor of collision term
 in the Boltzmann equation is given by
\begin{eqnarray}
\frac{s \langle\sigma v\rangle}{xH}
\simeq \frac{\lambda}{x_t^2} \langle\sigma v\rangle
 \label{collision-term}
\end{eqnarray}
with a temperature independent constant $\lambda$ defined as
\begin{eqnarray}
\frac{s}{xH}
= \frac{s}{x\sqrt{\frac{8\pi G}{3}\rho\left(1+\frac{\rho}{\rho_0}\right)}}
= \lambda\frac{x^{-2}}{\sqrt{1+ \left(\frac{x_t}{x} \right)^4}}
\label{xt}
\end{eqnarray}
where a new temperature independent parameter $x_t$ is defined as
 $x_t^4 =\frac{\rho}{\rho_0}\left. \right|_{T=m}$.
Note that the evolution of the universe can be divided into two eras.
At the era $x \ll x_t$ the $\rho^2$ term
 in Eq.~(\ref{BraneFriedmannEq}) dominates
 (brane world cosmology era),
 while at the era $x \gg x_t$ the expansion law
 obeys the standard cosmological law (standard cosmology era).
In the following, we call the temperature defined
 as $T_t = m x_t^{-1}$ (or $x_t$ itself) ``transition temperature''
 at which the evolution of the universe changes
 from the brane world cosmology era to the standard cosmology era.
We consider the case that
 the decoupling temperature of the dark matter particle
 is higher than the transition temperature.
In such a case, we can expect a considerable modification
 for the relic density of the dark matter
 from the one in the standard cosmology.

At the brane world cosmology era the Boltzmann equation
can be read as
\begin{equation}
\frac{d Y}{d x} = - \frac{\lambda}{x_t^2}
  \sigma_n x^{-n} (Y^2-Y_{EQ}^2).
\end{equation}
Here, for simplicity, we have parameterized
 the average of the annihilation cross section  
 times the relative velocity
 as $\langle \sigma v \rangle = \sigma_n x^{-n}$
 with a (mass dimension 2) constant $\sigma_n$.
Note that, in the right-hand side of the above equation,
 $x^2$ in the standard cosmology is replaced
 by the constant $x_t^2$.
At the early time, the dark matter particle is
 in the thermal equilibrium and $Y$ tracks $Y_{EQ}$ closely.
To begin, consider the small deviation from the thermal distribution
 $\Delta =Y-Y_{EQ} \ll Y_{EQ}$.
The Boltzmann equation leads to
\begin{eqnarray}
\Delta \simeq -\frac{x_t^2 \frac{d Y_{EQ}}{d x}}
  {\lambda \sigma_n x^{-n} (2Y_{EQ}+\Delta )}
\simeq \frac{x_t^2}{2\lambda \sigma_n x^{-n}},
\end{eqnarray}
where we have used an approximation formula
 $Y_{EQ}=0.145 (g/g_{*S}) x^{3/2} e^{-x}$ and
 $d Y_{EQ}/dx \simeq - Y_{EQ}$.
As the temperature decreases or equivalently $x$ becomes large,
 the deviation relatively grows
 since $Y_{EQ}$ is exponentially dumping.
Eventually the decoupling occurs at $x_d$
 roughly evaluated as $\Delta(x_d) \simeq Y(x_d) \simeq Y_{EQ}(x_d)$,
or explicitly
\begin{eqnarray}
\left. \frac{x_t^2}{2\lambda \sigma_n x^{-n}} \right|_{x=x_d}
\simeq \left. 0.145\frac{g}{g_{*S}}x^{3/2}e^{-x} \right|_{x=x_d}.
\end{eqnarray}
At further low temperature, $\Delta \simeq Y \gg Y_{EQ}$ is satisfied
 and $Y_{EQ}^2$ term in the Boltzmann equation can be neglected
 so that
\begin{equation}
\frac{d \Delta}{d x} = -
  \frac{\lambda}{x_t^2} \sigma_n x^{-n} \Delta^2.
\end{equation}
The solution is formally given by
\begin{eqnarray}
\frac{1}{\Delta(x)}-\frac{1}{\Delta(x_d)} =
\int_{x_d}^{x} \lambda \frac{\sigma_n x^{-n}}{x_t^2}
\label{sol}
\end{eqnarray}
For $n=0$ (S-wave process), $n=1$ (P-wave process), and $n >1$
we find   
\begin{eqnarray}  
  && \frac{\lambda \sigma_0}{x_t^2} \left( x - x_d \right),
   \nonumber \\
\frac{1}{\Delta(x)}-\frac{1}{\Delta(x_d)} =
 &&  \frac{\lambda \sigma_1}{x_t^2} \ln\left( \frac{x}{x_d} \right),
  \nonumber \\
 &&  \frac{\lambda \sigma_n}{x_t^2}
   \left( \frac{1}{n-1}\right)
\left( \frac{1}{x_d^{n-1}} - \frac{1}{x^{n-1}}
\right),
 \label{solutions}
\end{eqnarray}  
respectively.
Note that $\Delta(x)^{-1}$ is continuously growing 
 without saturation for $n \leq 1$. 
This is a very characteristic behavior 
 of the brane world cosmology, 
 comparing the case in the standard cosmology 
 where $\Delta(x)$ saturates after decoupling 
 and the resultant relic density is roughly 
 given by $Y(\infty) \simeq Y(x_d)$. 
For a large $x \gg x_d$ in Eq.~(\ref{solutions}), 
 $\Delta(x_d)$ and $x_d$ can be neglected. 
When $x$ becomes large further and reaches $x_t$, 
 the expansion law changes into the standard one, 
 and then $Y$ obeys the Boltzmann equation 
 with the standard expansion law for $x \geq x_t$. 
Since the transition temperature is smaller 
 than the decoupling temperature in the standard cosmology 
 (which case we are interested in), 
 we can expect that the number density freezes out 
 as soon as the expansion law changes into the standard one. 
Therefore the resultant relic density can be roughly 
 evaluated as $Y(\infty) \simeq \Delta(x_t)$ 
 in Eq.~(\ref{solutions}). 
In the following analysis, we will show that 
 this expectation is in fact correct. 

Now, we derive analytic formulas of the final relic density
 of the dark matter in the brane world cosmology.
At low temperature where $\Delta \simeq Y \gg Y_{EQ}$ is satisfied,
 the Boltzmann equation is given by
\begin{equation}
 \frac{d \Delta}{d x}
= - \frac{\lambda}{\sqrt{x^4+x_t^4}} (\sigma_n x^{-n}) \Delta^2 ,
\end{equation}
and the solution is formally described as
\begin{equation}
\frac{1}{\Delta(x)}-\frac{1}{\Delta(x_d)} =
\lambda \sigma_n  \int_{x_d}^{x} dy
 \frac{y^{-n}} {\sqrt{y^4+x_t^4}}.
\label{IntEqDelta}
\end{equation}
For $n=0$, the integration is given by the elliptic integral
 of the first kind such as
\begin{eqnarray}
\int_{x_d}^{x}dy \frac{1}{\sqrt{y^4+x_t^4}}
&=&
 \int^{x}_0 dy \frac{1}{\sqrt{y^4+x_t^4}}
 -\int^{x_d}_0 dy \frac{1}{\sqrt{y^4+x_t^4}} \nonumber \\
&=& \frac{2-\sqrt{2}}{x_t} \left[F\left(\arctan
(1+\sqrt{2})\frac{x_t+x}{x_t-x},
2\sqrt{3\sqrt{2}-4}\right)\right. \nonumber\\
 &-& \left.  F\left(\arctan (1+\sqrt{2})\frac{x_t+x_d}{x_t-x_d},
 2\sqrt{3\sqrt{2}-4}\right)\right] ,
\end{eqnarray}
where the elliptic integral of the first kind $F(\phi, k)$
is defined as
\begin{equation}
F(\phi, k) = \int^{\phi}_0\frac{d\theta}{\sqrt{1-k^2\sin^2\theta}} .
\end{equation}
In the limit $x \rightarrow \infty$,
we obtain (with appropriate choice of the phase $\phi$ in $F(\phi,k)$)
\begin{eqnarray}
\int_{x_d}^{\infty} \frac{d y}{\sqrt{y^4+x_t^4}}
&=& \frac{2-\sqrt{2}}{x_t}\left[
-F\left(\arctan (1+\sqrt{2})\frac{x_t+x_d}{x_t-x_d},
2\sqrt{3\sqrt{2}-4}\right)
\right. \nonumber \\
&+& \left.
 F \left(\pi-\arctan (1+\sqrt{2}),2\sqrt{3\sqrt{2}-4}\right)\right].
\end{eqnarray}
In the case of $x_d \ll x_t$,
 the integration gives $\simeq 1.85/x_t$, and
 we find the resultant relic density
 $Y(\infty) \simeq 0.54 x_t/(\lambda \sigma_0)$.
Note that the density is characterized
 by the transition temperature $x_t$ as we expected.
By using the well known formula (\ref{Ystandard}),
 for a given $\langle\sigma v\rangle$,
 we obtain the ratio of the energy density of the dark matter
 in the brane world cosmology ($\Omega_{(b)}$)
 to the one in the standard cosmology ($\Omega_{(s)}$)
 such that
\begin{eqnarray}
\frac{\Omega_{(b)}}{\Omega_{(s)}}
\simeq 0.54 \left(\frac{x_t}{x_{d(s)}}\right) ,
\end{eqnarray}
where $x_{d(s)}$ is the decoupling temperature
 in the standard cosmology.
Similarly, for $n=1$ we find
\begin{eqnarray}
\int_{x_d}^{\infty} dy \frac{y^{-1}}{\sqrt{y^4+x_t^4}}
= \left.\frac{1}{4x_t^2}\ln\left(\frac{\sqrt{x^4+x_t^4}-x_t^2}
{\sqrt{x^4+x_t^4}+x_t^2}\right)\right|^{\infty}_{x_d}
\simeq
-\frac{1}{4x_t^2}\ln\frac{\sqrt{x_d^4+x_t^4}-x_t^2}
{\sqrt{x_d^4+x_t^4}+x_t^2}.
\end{eqnarray}
Again, in the case of $x_d \ll x_t$,
 the integration gives 
 $\simeq x_t^{-2} \ln(x_t/x_d) \simeq x_t^{-2} \ln x_t$,
 and we find the resultant relic density
 $Y(\infty) \simeq x_t^2/(\lambda \sigma_0 \ln x_t)$.
Thus the ratio of the energy density of the dark matter
 is found to be
\begin{eqnarray}
\frac{\Omega_{(b)}}{\Omega_{(s)}}
\simeq \frac{1}{2 \ln x_t}
 \left(\frac{x_t}{x_{d(s)}}\right)^2 .
\end{eqnarray}
We can obtain results for the case of $n >1$
 in the same manner.  
Now we have found that the relic energy density
 in the brane world cosmology is characterized by
 the transition temperature and can be enhanced
 compared to the one in the standard cosmology,
 if the transition temperature is lower than
 the decoupling temperature.

In summary, we have investigated the thermal relic density
 of the cold dark matter in the brane world cosmology.
If the five dimensional Planck mass is small enough,
 the $\rho^2$ term in the modified Friedmann equation
 can be effective when the dark matter is decoupling.
We have derived the analytic formulas for the relic density
 and found that the resultant relic density can be enhanced.
The enhancement factor is characterized
 by the transition temperature, at which
 the evolution of the universe changes from the brane world cosmology era
 to the standard cosmology era.

It would be interesting to apply our result 
 to detailed numerical analysis 
 of the relic abundance of supersymmetric dark matter,
 for example, the neutralino dark matter.
Allowed regions obtained in the previous analysis
 in the standard cosmology \cite{CDM}
 would be dramatically modified
 if the transition temperature is small enough \cite{NOS}.
Furthermore, if there exists the brane world cosmology era
 in the history of the universe,
 the modified expansion law affects many physics
 controlled by the Boltzmann equations in the early universe.
For example, we can expect considerable modifications for
 the thermal production of gravitino \cite{OS2},
 the thermal leptogenesis scenario \cite{OS3} etc.
Those are worth investigating.

%
We would like to thank Takeshi Nihei for useful discussions. 
N.O. is supported in part by the Grant-in-Aid for 
Scientific Research  (\#15740164). 
O.S. is supported by the National Science Council of Taiwan 
under the grant No. NSC 92-2811-M-009-018.



\end{document}